\documentclass[unsortedaddress,superscriptaddress,prmaterials,twocolumn]{revtex4-2}
\usepackage{hyperref}
\usepackage{epsfig}
\usepackage{graphicx}
\usepackage{subfigure}
\usepackage{latexsym}
\usepackage{color}
\usepackage{fullpage}
\usepackage{dcolumn}
\usepackage{bm}
\usepackage[normalem]{ulem}
\usepackage{units}
\usepackage{amsmath}
\usepackage[paperwidth=210mm,paperheight=297mm,centering,hmargin=2cm,vmargin=2.3cm]{geometry}

\begin{document}


\title{Laser patterning of the room temperature van der Waals ferromagnet 1$T$-CrTe$_2$}

\author{Tristan Riccardi}
\thanks{These two authors contributed equally}
\affiliation{Univ. Grenoble Alpes, CNRS, Grenoble INP, Institut NEEL, 38000 Grenoble, France}
\affiliation{LNCMI-EMFL, CNRS UPR3228, Univ. Grenoble Alpes, Univ. Toulouse, Univ. Toulouse 3, INSA-T, Grenoble and Toulouse, France}
\author{Suman Sarkar}
\thanks{These two authors contributed equally}
\affiliation{Univ. Grenoble Alpes, CNRS, Grenoble INP, Institut NEEL, 38000 Grenoble, France}
\author{Anike Purbawati}
\affiliation{Univ. Grenoble Alpes, CNRS, Grenoble INP, Institut NEEL, 38000 Grenoble, France}

\author{Aloïs Arrighi}
\affiliation{Univ. Grenoble Alpes, CNRS, Grenoble INP, Institut NEEL, 38000 Grenoble, France}
\author{Marek Kostka}
\affiliation{Univ. Grenoble Alpes, CNRS, Grenoble INP, Institut NEEL, 38000 Grenoble, France}
\affiliation{Institute of Physical Engineering, Brno University of Technology, Brno 616 69, Czech Republic}

\author{Abdellali Hadj-Azzem}
\affiliation{Univ. Grenoble Alpes, CNRS, Grenoble INP, Institut NEEL, 38000 Grenoble, France}
\author{Jan Vogel}
\affiliation{Univ. Grenoble Alpes, CNRS, Grenoble INP, Institut NEEL, 38000 Grenoble, France}
\author{Julien Renard}
\affiliation{Univ. Grenoble Alpes, CNRS, Grenoble INP, Institut NEEL, 38000 Grenoble, France}
\author{La\"{e}titia Marty}
\affiliation{Univ. Grenoble Alpes, CNRS, Grenoble INP, Institut NEEL, 38000 Grenoble, France}

\author{Amit Pawbake}
\affiliation{LNCMI-EMFL, CNRS UPR3228, Univ. Grenoble Alpes, Univ. Toulouse, Univ. Toulouse 3, INSA-T, Grenoble and Toulouse, France}
\author{Clément Faugeras}
\affiliation{LNCMI-EMFL, CNRS UPR3228, Univ. Grenoble Alpes, Univ. Toulouse, Univ. Toulouse 3, INSA-T, Grenoble and Toulouse, France}

\author{Kenji Watanabe}
\affiliation{Research Center for Functional Materials, National Institute for Materials Science, 1-1 Namiki, Tsukuba 305-0044, Japan}
\author{Takashi Taniguchi}
\affiliation{International Center for Materials Nanoarchitectonics, National Institute for Materials Science, 1-1 Namiki, Tsukuba 305-0044, Japan}

\author{Aurore Finco}
\affiliation{Laboratoire Charles Coulomb, Universit\'{e} de Montpellier and CNRS, 34095 Montpellier, France}
\author{Vincent Jacques}
\affiliation{Laboratoire Charles Coulomb, Universit\'{e} de Montpellier and CNRS, 34095 Montpellier, France}

\author{Lei Ren}
\affiliation{Universit\'{e} de Toulouse, INSA-CNRS-UPS, LPCNO, 135 Av. Rangueil, 31077 Toulouse, France}
\author{Xavier Marie}
\affiliation{Universit\'{e} de Toulouse, INSA-CNRS-UPS, LPCNO, 135 Av. Rangueil, 31077 Toulouse, France}
\author{Cedric Robert}
\affiliation{Universit\'{e} de Toulouse, INSA-CNRS-UPS, LPCNO, 135 Av. Rangueil, 31077 Toulouse, France}

\author{Manuel Nu\~{n}ez-Regueiro}
\affiliation{Univ. Grenoble Alpes, CNRS, Grenoble INP, Institut NEEL, 38000 Grenoble, France}
\author{Nicolas Rougemaille}
\affiliation{Univ. Grenoble Alpes, CNRS, Grenoble INP, Institut NEEL, 38000 Grenoble, France}
\author{Nedjma Bendiab}
\affiliation{Univ. Grenoble Alpes, CNRS, Grenoble INP, Institut NEEL, 38000 Grenoble, France}
\author{Johann Coraux}
\email{johann.coraux@neel.cnrs.fr}
\affiliation{Univ. Grenoble Alpes, CNRS, Grenoble INP, Institut NEEL, 38000 Grenoble, France}

\begin{abstract}
Lamellar crystalline materials, whose layers are bond by van der Waals forces, can be stacked to form ultrathin artificial heterostructures, and in particular vertical magnetic junctions when some of the stacked materials are (ferro)magnetic. Here, using the room temperature van der Waals ferromagnet 1$T$-CrTe$_2$, we report a method for patterning lateral magnetic junctions. Exploiting the heat-induced phase transformation of the material into Cr$_x$Te$_y$ compounds ($x/y>1/2$), we use local laser heating to imprint patterns at the micron-scale. Optimizing laser heat dissipation, we further demonstrate the crucial role of the substrate to control the phase transformation. If plain, unstructured poorly heat-conducting substrates allow for direct writing of magnetic patterns, structured $h$-BN layers can serve as heat stencils to draw potentially thinner patterns. Besides, $h$-BN encapsulation turns out to be heat-protective (in addition from protecting against oxidation as it is generally used for), allowing the demonstration of room temperature ferromagnetism in $<$7~nm-thick 1$T$-CrTe$_2$.
\end{abstract}

\maketitle

\begin{figure}[!hbt]
\begin{center}
\includegraphics[width=78.4mm]{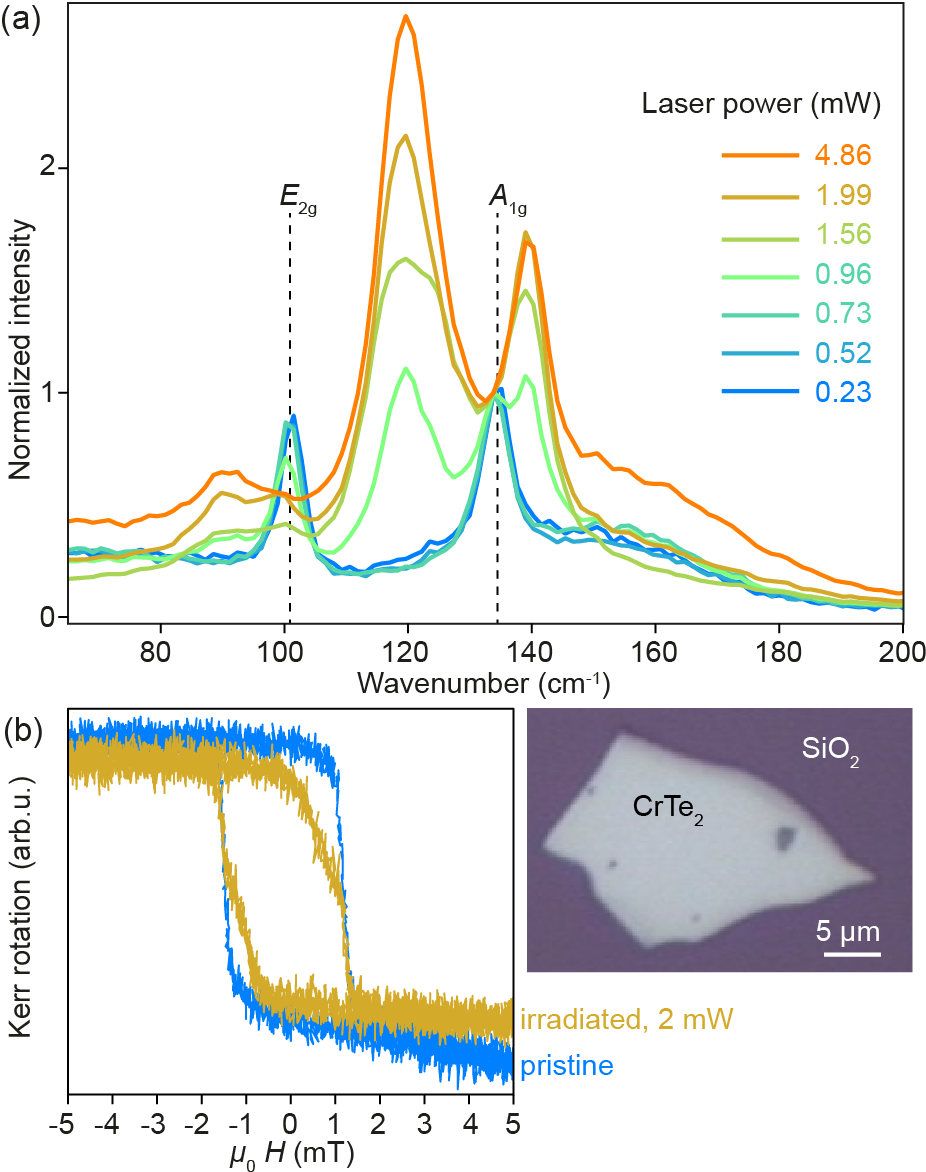}
\caption{\label{fig1} Laser-induced local transformation of 1$T$-CrTe$_2$ flakes on SiO$_2$/Si. (a) Raman spectra measured on a 30~nm-thick flake with increasing laser power (532~nm wavelength, acquisition time decreasing from 180~s to 30~s as power increases). (b) Focused-Kerr magnetometry (magnetic field applied in-plane), measured on a pristine (\textit{i.e.} only exposed to a very low laser fluence) and a laser-irradiated (still moderate 2~mW power) region of the same 1$T$-CrTe$_2$ flake (80~nm-thickness, optical micrograph shown on the right).}
\end{center}
\end{figure}

\textit{\textbf{Introduction. -- }}
The discovery of antiferromagnetic \cite{Lee,Wang} and ferromagnetic \cite{Huang,Gong} phases in two-dimensional (2D) crystals had considerable echo recently, setting large communities on track to revisit low-dimensional magnetic ordering and spintronic phenomena. Magnetic 2D materials give access to a wealth of fascinating properties, such as optimum electric-field control of magnetism, multifunctionalities, moir\'{e} effects, and flexible membrane-like architectures \cite{Gibertini,Wang_rev}. In that context, 2D materials exhibiting above room-temperature magnetism are of particular interest. However, such 2D compounds are rare and only very few \textit{bulk} van der Waals materials, from which 2D flakes may be exfoliated, have a Curie temperature $T_\mathrm{c}>300$~K. The 1$T$ phase of the CrTe$_2$ transition metal dichalcogenide \cite{Purbawati,Sun,Roeseler} with its colossal anomalous Hall effect \cite{Huang_b}, Fe$_5$GeTe$_2$ \cite{Chen} or Fe$_3$GaTe$_2$ \cite{Hu}, in both their bulk form and as exfoliated ultra-thin layers, are notable examples.

So far, van der Waals heterostructures essentially consist of vertical junctions of thin (or even purely 2D) materials \cite{Wang,Song,Yang,Shi}. Their functionality relies on an ultimately sharp interface between the layers, which are separated by a van der Waals gap. Additionally or alternatively, in-plane structuration can be introduced to pattern lateral junctions. Such lateral junctions offer further degrees of freedom for micro/nanostructuration and are suited for implementing or detecting various spin-dependent effects or functions, such as giant magnetoresistance, large spin accumulation or long-distance spin transport, nonlocal measurements of (inverse) spin Hall effect \cite{Jedema,Kimura,Tombros,Valenzuela,Ahn}. Lateral patterning can be achieved during the synthesis, by tuning growth conditions, as recently shown  with Cr-Te compounds \cite{Li}. Planar modulation of the electronic properties was demonstrated accordingly \cite{Li}. Rather than a bottom-up approach, a top-down strategy is also possible in principle. It however requires to locally transform the material, for instance with the help of a structural phase engineering under a focused laser beam. This strategy already allowed patterning ohmic junctions in a transition metal dichalcogenide \cite{Cho}, but was not used yet to our knowledge in the prospect of magnetic lateral junctions with van der Waals materials.

Here, we implement this laser-patterning approach using thin flakes of the room-temperature ferromagnet 1$T$-CrTe$_2$. In certain conditions, this material readily transforms into other Cr-Te compounds that are self-intercalated with Cr \cite{Saha,Purbawati_b,Figueiredo} and whose $T_\mathrm{c}$ falls below 300~K \cite{Purbawati_b}. This tendency to polymorphism explains why chemical vapor deposition of Cr-Te flakes can naturally produce (magnetic) Cr$_2$Te$_3$/Cr$_5$Te$_8$ lateral junctions \cite{Niu}, and the reason why moderate heating of 1$T$-CrTe$_2$ transforms it into other Cr-Te compounds (especially, Cr$_5$Te$_8$) \cite{Purbawati_b}. Using specific peaks in Raman scattering spectra as a thermometer and focused-Kerr magnetometry we show that (i) laser-induced thermal transformation of 1$T$-CrTe$_2$ can be strongly mitigated, and thus controlled, using an appropriate substrate, and (ii) patterns can be laser-imprinted within a ferromagnetic matrix, with a spatial resolution set by the optical beam, or with a potentially better resolution, limited by our ability to pattern a heat-dissipation material (e.g. hexagonal boron nitride, $h$-BN) in contact with the CrTe$_2$ flakes. We finally find that (iii) full $h$-BN encapsulation, usually prescribed to limit the physico-chemical interaction of 2D materials with airborne species, is also valuable to protect ultra-thin flakes that are particularly susceptible to laser-induced heating already at low laser powers, and thereby demonstrate room temperature ferromagnetism for a 1$T$-CrTe$_2$ thickness below 7~nm.

\textit{\textbf{Methods -- }}
The 1$T$-CrTe$_2$ bulk crystals (typically few 1~mm-wide and few 100~$\mu$m-thick) were synthesized using an elemental mixture of K, Cr and Te, introduced inside an evacuated quartz tube then heated to 1,170~K for eight days, and slowly cooled-down \cite{Freitas}. Potassium deintercalation was achieved using an iodine solution in acetonitrile, and the iodine was subsequently removed by washing with acetonitrile. The product was finally filtered and dried under vacuum. From the resulting 1$T$-CrTe$_2$ macroscopic platelets, thin flakes were exfoliated using scotch tape. Note that the ultimate 2D limit, of one- or few-layer flakes is so far not accessible (our thinnest exfoliated 1$T$-CrTe$_2$ flakes with measured room-temperature ferromagnetism comprise 11 layers, see below). Exfoliation within an argon-filled glove-box or in air does not change the flakes' properties, ruling out significant alteration upon exposure to air at the hour-timescale considered here.

The flakes were transferred onto three different substrates: (i) 285~nm-thick and 85~nm-thick SiO$_2$ on Si, (ii) Pt/Ta films (1~nm-thick Pt deposited on 10~nm-thick Ta by sputter-deposition on Si) and (iii) $h$-BN flakes using deterministic dry-transfer based on polydimethylsiloxane stamps, monitored under the objective of an optical microscope \cite{Castellanos,Dean}. The flakes' thickness was measured with a Bruker Dimension Icon atomic force microscope (AFM). The samples were heated under vacuum ($<$10$^{-3}$~mbar) using a variable-temperature Linkam HFS350EV-PB4 stage, and locally with a focused laser (see next paragraph).

Raman spectroscopy and local heating were performed with two experimental setups. The first one consists of a Witec Alpha 500 Raman microscope, a 532~nm laser focused through a $\times$50 objective (Mitutoyo, 0.75 numerical aperture) to a 1~$\mu$m spot, a Rayshield coupler and a 1,800~lines/mm grating (spectral resolution $\leq$0.1~cm$^{-1}$). The second setup, with which we explored a broader temperature range than addressed in the main text, uses a 515~nm excitation laser focused with a $\times$50 long working distance objective on a $\sim$1~$\mu$m spot. The sample is mounted on the cold finger of a helium flow cryostat. The scattered signal is analyzed by a 50~cm spectrometer equipped with a liquid-nitrogen-cooled charge coupled device camera. Measurements were done on samples sealed within Ar-filled quartz cells, in vacuum and in air, with no noticeable difference on the results.

Focused-Kerr magnetometry was performed with a home-made setup including a He-Ne laser (632~nm), a $\times$100 objective producing a 1~$\mu$m-wide, $s$-polarized spot on the sample surface, a Wollaston prism splitting the reflecting beam in two beams with orthogonal polarizations and analyzed by two identical photodiodes. The sum of the two corresponding signals was acquired during sweeps (1-2~Hz frequency, 100 sweeps typically used to obtain satisfactory signal-to-noise ratio) of an external magnetic field, applied within the sample surface \textit{via} a small horseshoe electromagnet.

\textit{\textbf{Laser-induced local transformations -- }}
Figure~\ref{fig1}a shows a series of Raman scattering spectra acquired at increasing laser power on the same location of a 1$T$-CrTe$_2$ flake (30~nm-thick) placed onto a SiO$_2$/Si substrate \footnote{The (metallic) flakes are thick-enough to be fully reflective, so interference effects that would involve light reflection at the SiO$2$/Si interface underneath are irrelevant here. We hence do not expect the effects we report to depend on the oxide thickness, as was the case for few layer graphene \cite{Han}.}. As long as the power density does not exceed 0.73~mW/$\mu$m$^2$, the spectra look qualitatively the same (we will come back to faint differences later). Two peaks are observed around 102~cm$^{-1}$ and 135~cm$^{-1}$, corresponding to the characteristic vibrational modes \cite{Purbawati} of 1$T$-CrTe$_2$ ($E_\mathrm{2g}$, $A_\mathrm{1g}$). At higher power, other contributions are observed in the spectra, around 125~cm$^{-1}$ and 144~cm$^{-1}$. They do not disappear after reducing the power below a threshold of about 0.75~mW (see Fig.~S1 in the Supplemental Material \cite{SM}), indicating a non-reversible transformation of the material. Such a spectroscopic signature is reminiscent of those of Cr$_x$Te$_y$ compounds with $x/y$ ratios beyond 1/2 \cite{Fu,Chen_b,Zhong,Yang_b,Gowda,Ghorai}. These peaks do not relate to oxidation of the material (test experiments have been performed with samples sealed within an Ar-filled quartz cell), but are precisely those appearing as 1$T$-CrTe$_2$ transforms into other Cr-Te phases upon heating (as established by some of us, annealing CrTe$_2$ flakes on a hot plate in a controlled atmosphere and carefully monitoring the concomitant compositional, structural, vibrational, and magnetic changes, see Ref.~\cite{Purbawati_b}). Transformation to Cr$_5$Te$_8$ is complete after 800~K annealing, and already starts at 500~K, with no CrTe$_2$ left at intermediate temperatures \cite{Purbawati_b}. It is also known that focused laser beams can locally heat 2D materials \cite{Balandin,Sahoo} and in some case even modify their structure \cite{Cho}; this is a straightforward interpretation for our observation too.

The changes observed with Raman scattering spectroscopy are expected to be concomitant with changes in the magnetic properties, since the different Cr-Te compounds have different Curie temperatures, below 300~K for Cr:Te atomic ratios beyond 1/2 \cite{Saha,Purbawati_b}. This is confirmed by focused Kerr-effect magnetometry. Figure~\ref{fig1}b shows two hysteresis loops measured at 300~K with an in-plane applied magnetic field on a $\sim$80~nm-thick flake. The first loop has been acquired on a region of the flake that was never exposed to laser powers above 0.4~mW (and with very short exposure times, typically a few 1~s), while for the second loop a region heated with a 2~mW laser beam was spotted (laser power / temperature calibration experiments discussed below show that such power increases temperature by $\sim$170~K). The two measurements were performed in the same run, without changing the optical setup. The second loop clearly shows reduced Kerr rotation angle and less square loop, pointing to a smaller amount of ferromagnetic (1$T$-CrTe$_2$) material.

These data establish a partial phase transformation of CrTe$_2$ into other Cr-Te compounds, which are known to magnetically order well below 300~K, under laser-induced heating. While this can be exploited and optimized to write magnetic/non-magnetic patterns with a laser as we will discuss later, we anticipate that future devices based on CrTe$_2$-on-SiO$_2$ might be only moderately resilient to local Joule heating typically produced by the high current density in modern nanoscaled architectures.

\begin{figure}[!hbt]
\begin{center}
\includegraphics[width=68.4mm]{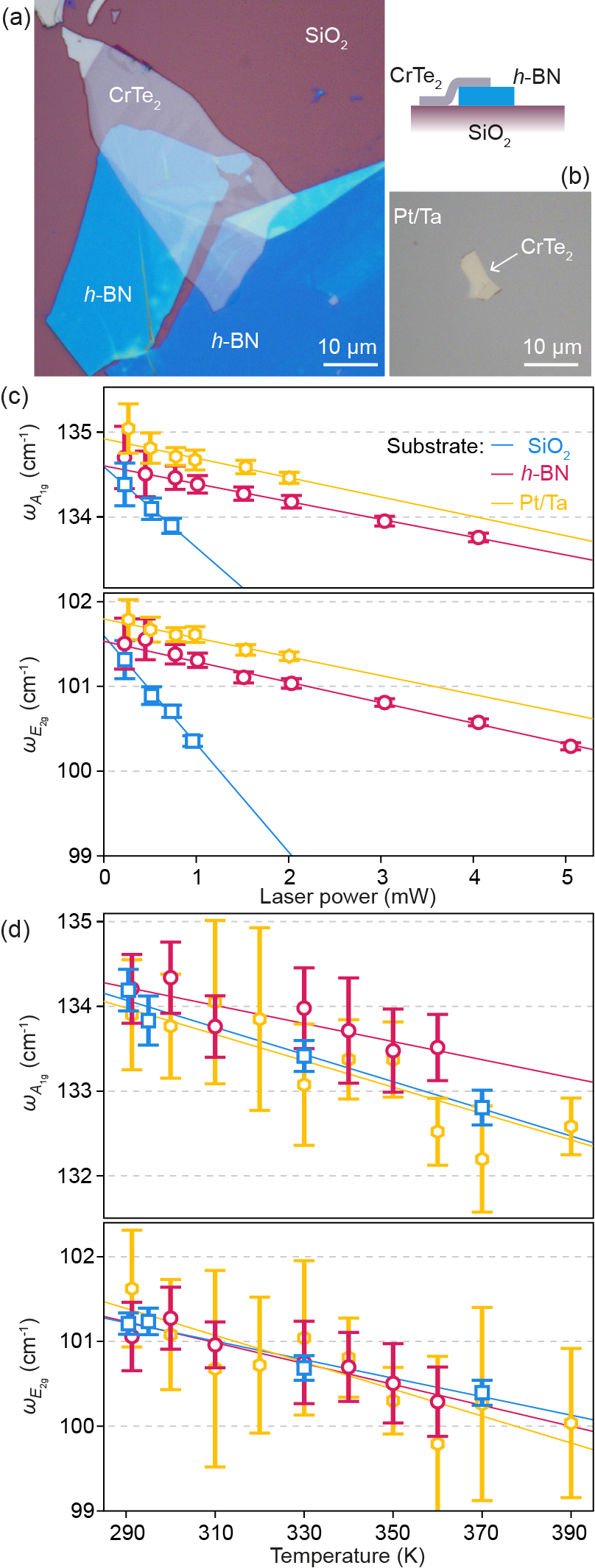}
\caption{\label{fig2} Raman thermometry of Cr-Te flakes on different substrates. (a,b) Optical images of the two samples, a stack consisting of a $\sim$30~nm-thick 1$T$-CrTe$_2$ flake on a 25~nm-thick $h$-BN buffer layer, transfered to a Si/SiO$_2$ substrate (a), and a 20~nm-thick 1$T$-CrTe$_2$ flake on a Pt/Ta substrate (b). (c,d) Variations of the central wavenumber of the $E_\mathrm{2g}$ and $A_\mathrm{1g}$ modes, $\omega_{E_\mathrm{2g}}$ and $\omega_{A_\mathrm{1g}}$, as function of the power of the laser beam (c) and the sample temperature (d), on the three substrates (Si/SiO$_2$, $h$-BN, Pt/Ta).}
\end{center}
\end{figure}

\textit{\textbf{Substrate-controlled heating  -- }}
Besides the above-discussed obvious qualitative changes in the Raman scattering spectra of 1$T$-CrTe$_2$, the central wavenumbers $\omega_{E_\mathrm{2g}}$ and $\omega_{A_\mathrm{1g}}$ of the two peaks corresponding to the $E_\mathrm{2g}$ and $A_\mathrm{1g}$ modes are quantitatively affected by the laser power.

\begin{table*}[!bt]
\caption{\label{tab:slopes}Variation of the wavenumber of the two Raman modes of 1$T$-CrTe$_2$, $\omega_{E_\mathrm{2g}}$ and $\omega_{A_\mathrm{1g}}$, with temperature $T$ ($\Delta\omega_{E_\mathrm{2g}}/\Delta T$, $\Delta\omega_{A_\mathrm{1g}}/\Delta T$) and with laser power $P$ ($\Delta\omega_{E_\mathrm{2g}}/\Delta P$, $\Delta\omega_{A_\mathrm{1g}}/\Delta P$), on different substrates.}
\begin{ruledtabular}
\begin{tabular}{ccccc}
Substrate & $\Delta\omega_{E_\mathrm{2g}}/\Delta T$ & $\Delta\omega_{E_\mathrm{2g}}/\Delta P$ & $\Delta\omega_{A_\mathrm{1g}}/\Delta T$ & $\Delta\omega_{A_\mathrm{1g}}/\Delta P$ \\[0.1cm]
\hline
SiO$_2$ & -0.011(2)~cm$^{-1}$K$^{-1}$ & -1.3(2)~cm$^{-1}$mW$^{-1}$ & -0.016(4)~cm$^{-1}$K$^{-1}$ & -0.9(5)~cm$^{-1}$mW$^{-1}$ \\
$h$-BN & -0.012(6)~cm$^{-1}$K$^{-1}$ & -0.24(1)~cm$^{-1}$mW$^{-1}$ & -0.011(7)~cm$^{-1}$K$^{-1}$ & -0.21(2)~cm$^{-1}$mW$^{-1}$ \\
Pt/Ta & -0.016(8)~cm$^{-1}$K$^{-1}$ & -0.22(6)~cm$^{-1}$mW$^{-1}$ & -0.016(5)~cm$^{-1}$K$^{-1}$ & -0.23(8)~cm$^{-1}$mW$^{-1}$ \\
\end{tabular}
\end{ruledtabular}
\end{table*}

Our analysis is made on three different substrates, namely the surface of SiO$_2$, a 11~nm-thick Pt/Ta layer on Si, and a 25~nm-thick $h$-BN layer (Figs.~\ref{fig2}a,b). In the following we address 1$T$-CrTe$_2$ flakes whose thickness is 30~nm at most, which we find is optimum for exploiting laser heating effects using reasonably accessible laser powers. After careful optimization of our measurement protocol, key for our conclusions was the ability to combine a full set of Raman scattering spectroscopy and focused Kerr magnetometry characterizations on the very same sample (SiO$_2$/Si partially covered with $h$-BN).

Figure~\ref{fig2}c reveals a linear red-shift of both $\omega_{E_\mathrm{2g}}$ and $\omega_{A_\mathrm{1g}}$ extracted from data such as shown in Fig.~\ref{fig1}a, on the three substrates, with increasing laser power. Such red-shifts of the lattice's vibrational modes upon heating, also observed in graphene \cite{Balandin} and MoS$_2$ \cite{Sahoo}, translate the anharmonicity of the interatomic potential.

Two main observations can be made. First, on a SiO$_2$ substrate, the modification of the Raman spectra are so severe above 1~mW that it becomes impossible to extract the $\omega_{E_\mathrm{2g}}$ and $\omega_{A_\mathrm{1g}}$ values, so that only three/four points have been reported on Fig.~\ref{fig2}c, unlike for the other substrates. Second, the slopes of $\omega_{E_\mathrm{2g}}$ and $\omega_{A_\mathrm{1g}}$ versus power are similar on the Pt/Ta and $h$-BN substrate ($\sim-0.2$~cm$^{-1}$mW$^{-1}$), and several times smaller than those on SiO$_2$ (Tab.~\ref{tab:slopes}). Note that the thickness of the flake (30~nm) is the same on SiO$_2$ and $h$-BN (the very same flake lays on both substrates), and that we selected a flake of similar thickness (20~nm) on the Pt/Ta substrate.

These observations indicate that the laser heating is far less prominent using a Pt/Ta or a $h$-BN substrate than using a SiO$_2$/Si substrate \footnote{No heat-sink effect is observed here with SiO$_2$/Si, unlike what was reported with few-layer graphene irradiated, on this substrate, with large laser powers (60~mW) \cite{Han}, or if this effect exists in our case, it appears to be far less prominent than on Pt/Ta or $h$-BN substrates.}. Heat dissipation is hence more efficient with the first two substrates. They are indeed good heat conductors, Pt/Ta obviously \textit{via} electrons as a metal, with thermal conductivity $\kappa\sim70$~W$\cdot$m$^{-1}$K$^{-1}$, and $h$-BN as well, \textit{via} phonons, with $\kappa\sim400$~W$\cdot$m$^{-1}$K$^{-1}$ \cite{Jo} in the range of thicknesses we address here. Actually, $h$-BN was also shown to stabilise MoS$_2$ during heat treatments, and to help dissipating heat in MoS$_2$/$h$-BN stacks \cite{Arrighi}. Unlike Pt/Ta and $h$-BN, SiO$_2$ is a poor heat conductor, with $\kappa\sim1$~W$\cdot$m$^{-1}$K$^{-1}$. 

\begin{figure*}[!bt]
\begin{minipage}[c]{0.623\linewidth}
\includegraphics[width=\linewidth]{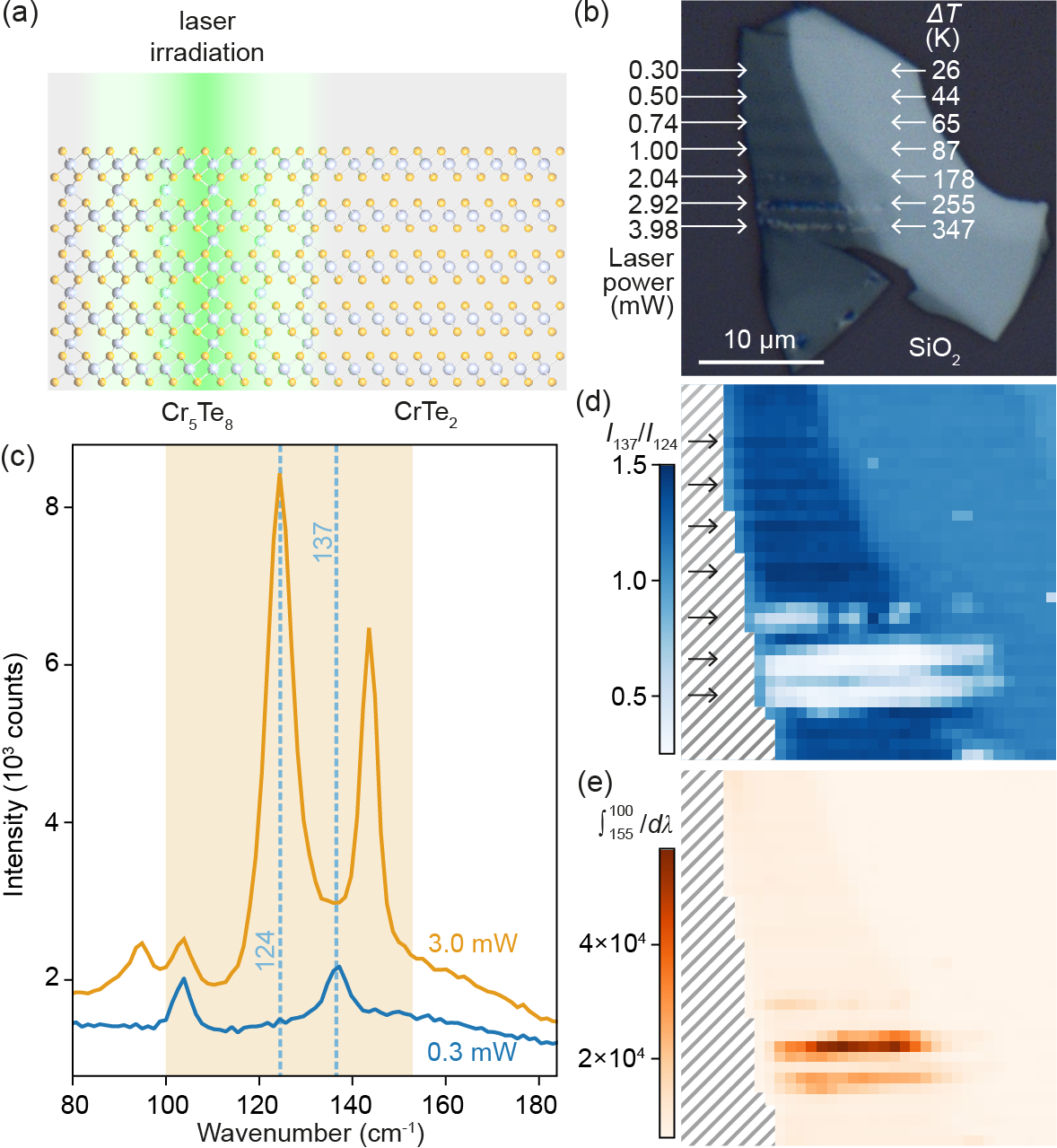}
\end{minipage}\hfill
\begin{minipage}[c]{0.33\textwidth}
\caption{\label{fig3} Direct laser patterning of 1$T$-CrTe$_2$. (a) Cross-section cartoon of the atomic structure representing the laser-induced tranformation of 1$T$-CrTe$_2$ into a Cr self-intercalated Cr-Te compound (here Cr$_5$Te$_8$). (b) Optical image of a flake on SiO$_2$/Si, laser-irradiated along line scans (whose extension is delimited with arrows) of increasing power. The laser-induced increases of temperature, $\Delta T$, are indicated. The flake has regions of different thicknesses; here we specifically consider the thinner region of darkest optical contrast, whose thickness is $\sim$20~nm. (c) Representative Raman spectra measured with a low (0.3~mW) and a high (3.0~mW) laser power. (d,e) Hyperspectral Raman mapping (0.3~mW laser power for acquisition): (d) Ratio of the intensity averaged around 137~cm$^{-1}$ and 124~cm$^{-1}$ (transformed 1$T$-CrTe$_2$ appears with deep hue of blue); (e) Area under the spectra between 100~cm$^{-1}$ and 155~cm$^{-1}$ (unaltered 1$T$-CrTe$_2$ appears with deep hue of orange).}
\end{minipage}
\end{figure*}

To establish a correspondance between the laser power and the corresponding local temperature, we performed a series of calibration experiments: the $\omega_{E_\mathrm{2g}}$ and $\omega_{A_\mathrm{1g}}$ red-shifts were studied as function of a controlled global temperature applied to the samples as a whole (Fig.~\ref{fig2}d, measured in vacuum) \footnote{Although our focus here is on the properties of the van der Waals ferromagnet around room temperature and above it (which may be of relevance for future practical applications), we report, as Supplemental Material Fig.~S2 $\omega_{E_\mathrm{2g}}(T)$ and $\omega_{A_\mathrm{1g}}(T)$ for temperature $T\in[4~\mathrm{K},300~\mathrm{K}]$ on bulk  1$T$-CrTe$_2$ flake \cite{SM}.}. In the 100~K temperature range we explored, we observe linear red-shifts for $\omega_{E_\mathrm{2g}}$ and $\omega_{A_\mathrm{1g}}$, this time with similar slopes on all three substrates, between $-0.011$~cm$^{-1}$K$^{-1}$ and $-0.016$~cm$^{-1}$K$^{-1}$, which seems reasonable as the flakes are expectedly thermalized on their substrates (the uncertainties on the determination of $\omega_{E_\mathrm{2g}}$ and $\omega_{A_\mathrm{1g}}$ are larger than for the laser-power-dependent measurements, due to the use of a vacuum cell that imposes a relatively large sample-to-objective distance altering the signal-to-noise ratio). Coming back to the laser-power-dependent data (Fig.~\ref{fig2}c), we can now tell that from 1~mW to 5~mW, the local temperature increases by less than 100~K on Pt/Ta and $h$-BN substrates, while it increases by about 400~K on SiO$_2$/Si. The latter is more than enough to fully transform 1$T$-CrTe$_2$, especially into Cr$_5$Te$_8$ \cite{Purbawati_b}.


Modeling heat transfer processes at the different interfaces between the flakes and their substrate, to quantitatively reproduce the local temperature increase, would require knowledge on several materials properties (e.g. out-of-plane thermal expansion coefficient in a van der Waals material / substrate stack \cite{Park}, thermal conductivities \cite{Yue}, which are anisotropic for 1$T$-CrTe$_2$ and other Cr$_x$Te$_y$ compounds, thermal impedances and phonon transmission \cite{Liu} at interfaces with different substrates, which depend on the interfaces' quality) that are so far ill-characterized (if at all). We hence leave this task for future works.

\textit{\textbf{Lateral magnetic patterning. -- }}
Properly choosing the substrate for the 1$T$-CrTe$_2$ flakes and using local laser irradiation, we can now locally heat the material to design planar junctions between a room temperature ferromagnet, 1$T$-CrTe$_2$, and another, non-magnetic Cr-Te compound (at least at 300~K). To do so, we will use laser power higher than 2~mW, for which the transformation of 1$T$-CrTe$_2$ is only partial (Fig.~\ref{fig1}b). Figure~S3 of the Supplemental Material shows magnetization \textit{versus} temperature for the bulk material, from which the flakes are exfoliated, before/after annealing at 600~K, the kind of temperature we target in the following laser-induced heating process \cite{SM}, and the concomitant lowering of the ordering temperature below 300~K.

A first approach consists in direct laser writing (Fig.~\ref{fig3}a): the focused ($\sim$1~$\mu$m) laser beam is scanned (0.03~$\mu$m/s) to draw traces on 1$T$-CrTe$_2$ flakes deposited on SiO$_2$/Si, which prevents efficient heat dissipation and allows reaching locally-high temperatures as we just saw. To illustrate this process, parallel lines were drawn on a flake (20~nm-thick), with increasing power from 0.50~mW to 3.98~mW (Fig.~\ref{fig3}b). According to our calibration of the temperature/power correspondance, laser-induced temperature increases $\Delta T$ from 44~K to 347~K ($\pm30\%$) are expected.

To monitor the effect of this laser patterning, we performed hyperspectral Raman imaging of the sample surface. Figure~\ref{fig3}c shows two representative Raman scattering spectra from such a hyperspectral dataset, acquired at two locations, where a flake has been irradiated with, respectively, low and high laser power. We extract two quantities. The first one, the ratio of intensities at 137~cm$^{-1}$ and 124~cm$^{-1}$, takes highest values where the probed material is 1$T$-CrTe$_2$. On the contrary, the second quantity, the scattered Raman intensity integrated from 100~cm$^{-1}$ to 155~cm$^{-1}$, is highest where the probed material has been fully transformed.

Maps of these two quantities are shown in Fig.~\ref{fig3}d,e. They look essentially complementary, and reveal that full transformation of the starting 1$T$-CrTe$_2$ is achieved with a laser power above 2~mW, of about 3~mW.

\begin{figure}[!ht]
\begin{center}
\includegraphics[width=80mm]{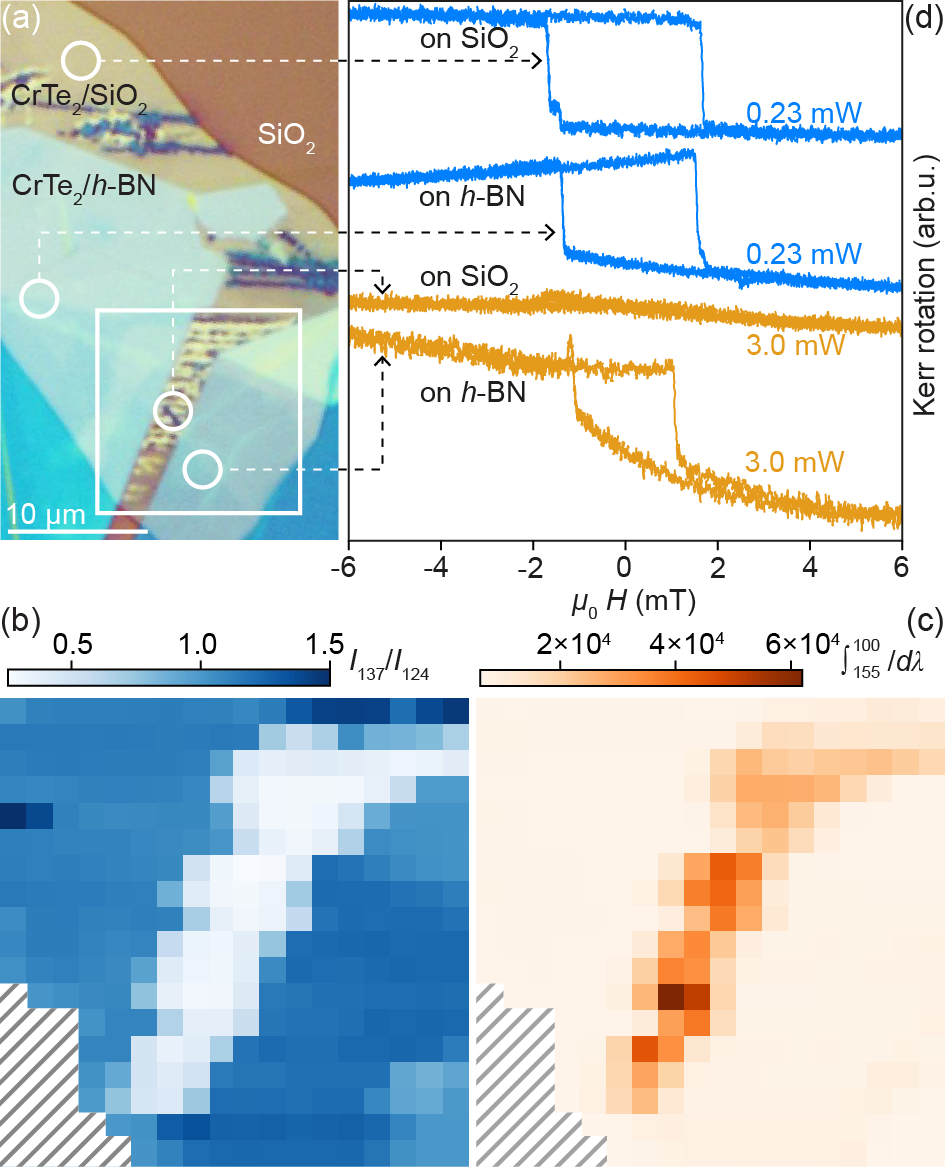}
\caption{\label{fig4} Laser patterning of 1$T$-CrTe$_2$ assisted with a structured, $h$-BN heat-stencil substrate.  (a) Optical micrograph of the same sample as in Fig.~\ref{fig2}a, after the sample has been raster-irradiated with a 3.0~mW laser within the region delimited with a white frame. Note the crack in the $h$-BN buffer, where 1$T$-CrTe$_2$ lays directly onto SiO$_2$. (b,c) Hyperspectral Raman mapping of the irradiated region: (b) Ratio of the scattered Raman intensity averaged around 137~cm$^{-1}$ and 124~cm$^{-1}$ (transformed 1$T$-CrTe$_2$ appears with deep hue of blue); (d) Area under the Raman scattering spectra between 100~cm$^{-1}$ and 155~cm$^{-1}$ (unaltered 1$T$-CrTe$_2$ appears with deep hue of orange). (d) Kerr rotation measured as function of the external in-plane magnetic field with a focused laser beam, on the region irradiated with a 3.0~mW laser and away from it (laser power not exceeding 0.23~mW). The data are acquired for sample regions laying on SiO$_2$ and $h$-BN.}
\end{center}
\end{figure}

The second approach is more elaborate, as it combines laser writing with structuration of the substrate, or rather here, insertion of a structured buffer layer ($h$-BN) between the substrate and 1$T$-CrTe$_2$. The buffer will allow to spatially modulate heat dissipation from the backside of the 1$T$-CrTe$_2$ flake. Setting the laser power above 2~mW, e.g. at 3~mW, temperature is expected to locally increase by a large $\Delta T\sim260$~K on SiO$_2$ and by only $\sim$60~K on $h$-BN.\footnote{This argument assumes patterns in $h$-BN with lateral dimensions large enough that a 1$T$-CrTe$_2$ flake can bend down to come in contact with the underlying substrate; one may also consider finer patterns, above which the flake will be suspended. There, heat dissipation will obviously marginally occur perpendicular to the flake and substantial heating may also occur.} Such a power seems a good compromise to selectively transform 1$T$-CrTe$_2$ areas defined by the $h$-BN flakes inserted between a SiO$_2$/Si substrate and the flake. In this sense, $h$-BN flakes would act as heat-dissipation masks (\textit{i.e.} heat stencils), and it is conceivable to pattern them with e.g. electron-beam lithography, to draw a variety of patterns. Here, as a proof of principle we use a more straightforwardly structured $h$-BN mask. Actually, manipulating the flakes in the process of stacking them sometimes generates cracks in the layers. Such an ill-controlled crack occurred in the $h$-BN layer that was selected to fabricate the stack shown in Fig.~\ref{fig2}a (the same sample is addressed in Fig.~\ref{fig4}). The sample then features the same 1$T$-CrTe$_2$ flake, laying partly onto $h$-BN, and partly onto SiO$_2$/Si at the location of the $\sim$2~$\mu$m-wide crack in $h$-BN.

Raster-scanning (0.07~$\mu$m/s) a 3~mW focused laser beam across a square area comprising this crack (Fig.~\ref{fig4}a), and performing hyperspectral imaging and analysis just like was done with the previous sample, the two complementary Fig.~\ref{fig4}b and Fig.~\ref{fig4}c prove here also a selective, local full transformation of 1$T$-CrTe$_2$.

Figure~\ref{fig4}d presents the focused-Kerr magnetometry analysis of the laser-irradiated region, compared to that of pristine 1$T$-CrTe$_2$. Away from the irradiated zone, the hysteresis loops on a SiO$_2$ or a $h$-BN substrate look similar, with a slight asymmetry compared to a pure square loop and a slightly reduced coercivity (compare the two top loops in Fig.~\ref{fig4}d) on $h$-BN, which can be due to lateral drift during the $\sim$10~min-long measurement and a locally slightly different thickness, respectively. In the irradiated zone, the presence of the $h$-BN layer between 1$T$-CrTe$_2$ and SiO$_2$ correlates with a profoundly different magnetic response: a somehow square loop is observed on $h$-BN (again, coercivity is different possibly due to a different thickness and the asymmetry with respect to a square loop is ascribed to lateral drift during data acquisition), and no magnetic signal is detected on SiO$_2$. This is fully consistent with the above Raman scattering analysis (performed on the very same location on the sample), altogether confirming the full transformation of 1$T$-CrTe$_2$ into other Cr-Te compound(s) (none exhibiting magnetic order at the 300~K temperature of the measurements) in absence of the buffer, heat-evacuating $h$-BN layer.

Note that this second approach is \textit{a priori} not limited by the size of the laser spot, so in principle finer patterns could be drawn with sub-optical-wavelength patterns etched into $h$-BN.

\textbf{\textit{Heat-protective capping with }h\textit{-BN. -- }}
Most studies of magnetism in few-layer or even single layers of Cr-Te compounds have been made with films grown by molecular beam epitaxy or chemical vapor deposition \cite{Meng,Chua,Zhang,Ou,Sun_b,Li_b,Lasek}. Obtaining such thin layers of Cr-Te compounds with exfoliation has so far been elusive (a noticeable exception is the ultrasonication of ferromagnetic CrTe bulk crystal into single-, bi-, and tri-layers of CrTe \cite{Wu}). As far as exfoliated 1$T$-CrTe$_2$ is concerned, the thinnest flakes reported to date have eight layers \cite{Sun}, and the thinnest flakes with room-temperature ferromagnetism have 14 layers \cite{Sun}, while most published data rather deal with $\geq$20~nm exfoliated films \cite{Sun,Purbawati,Purbawati_b,Huang_b}. Interestingly, no room temperature ferromagnetism could be observed in 16-layer 1$T$-CrTe$_2$ on a SiO$_2$/Si substrate \cite{Purbawati}, but the 14-layer 1$T$-CrTe$_2$ sample that showed room-temperature ferromagnetism was capped with a 5~nm Pt film \cite{Sun}. Overall, these laser-based magnetometry observations seem consistent with the possible laser-induced transformation of 1$T$-CrTe$_2$ we address in the present work, especially above a certain power threshold or when no capping material is used.

The transformation actually occurs more readily, at a given (moderate, 0.3~mW) laser power, for thin flakes than for thicker ones, as evident from Raman scattering spectroscopy (see Supplemental Material Fig.~S4 \cite{SM}, comparing the spectra of a 10~nm-thick and a 50~nm-thick flake). At this point it is important to stress that previous detailled chemical analysis by energy-dispersive X-ray spectroscopy performed by some of us ruled out the presence of oxygen (within the technique's sensitivity, an atomic fraction of few percent at most) in flakes exposed to air and laser-irradiated (0.2-0.4~mW with the same objective lens as in the present work) \cite{Purbawati_b}. From these considerations we can exclude oxidation of the flakes as an explanation for the observed changes in the Raman scattering spectra, and we instead propose that very thin flakes can only distribute the heat load over a small region of the crystal, and are thus more susceptible to laser-induced heating and phase transformation.

\begin{figure}[!ht]
\begin{center}
\includegraphics[width=80mm]{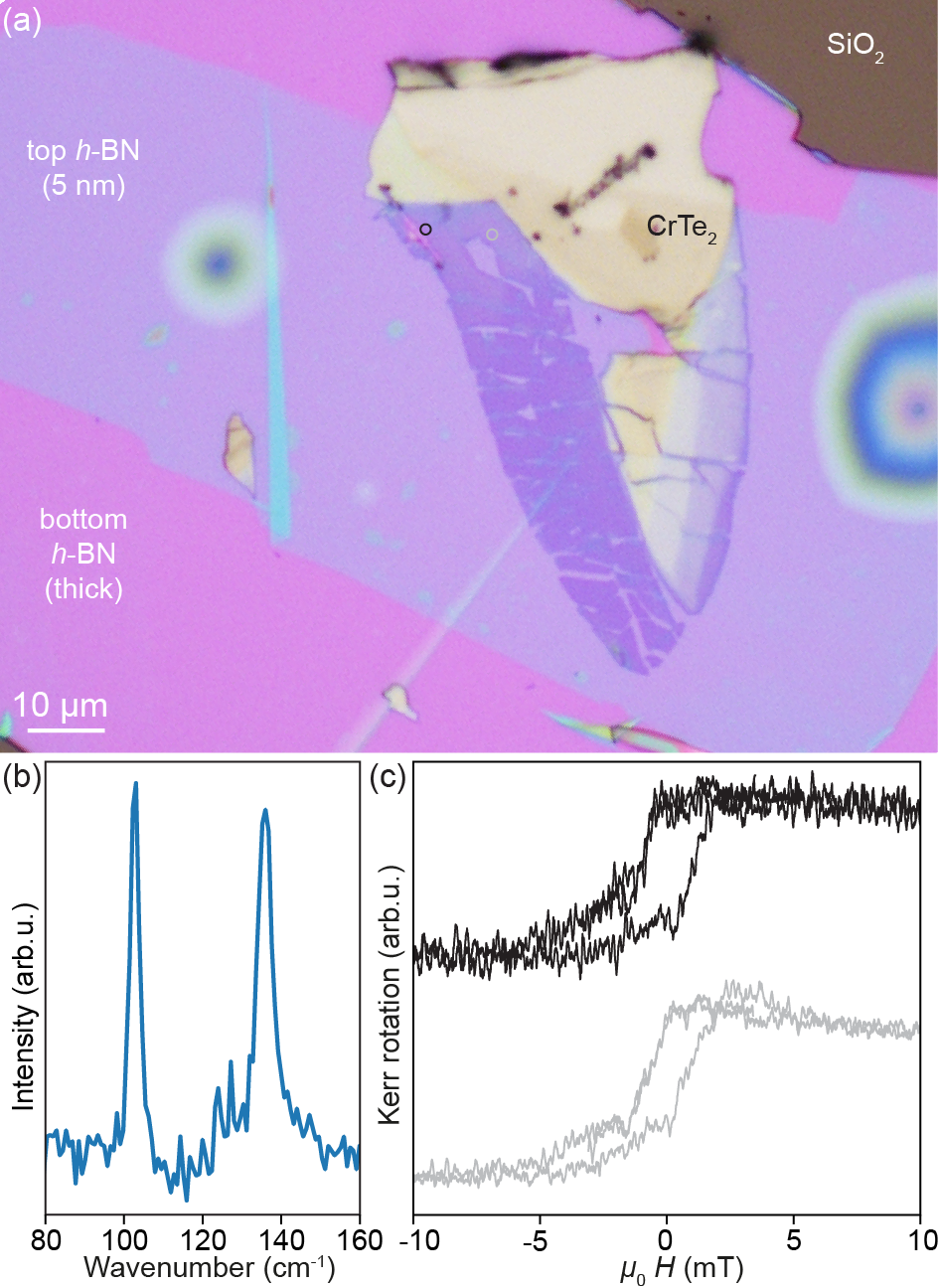}
\caption{\label{fig5} Thin encapsulated 1$T$-CrTe$_2$ with room-temperature ferromagnetism. (a) Optical image of a $h$-BN/1$T$-CrTe$_2$/$h$-BN sandwich, deposited on SiO$_2$/Si. The thickness of the 1$T$-CrTe$_2$ flake varies from $\sim$100~nm (beige regions) to 5.0~nm (dark pink contrast), with 6.9~nm-thick regions (less dark shade of pink, at the location of the green and orange circles). (b) Raman spectrum acquired for a 6.9~nm-thick 1$T$-CrTe$_2$. (c) Focused-Kerr data as function of applied in-plane magnetic field, measured at 300~K on two different locations of 6.9~nm-thick 1$T$-CrTe$_2$, with 0.17~mW [bottom loop, gray circle on (a)] and 0.25~mW [top loop, black circle on (a)] laser powers.}
\end{center}
\end{figure}

Supporting these ideas, our numerous attempts to detect room-temperature ferromagnetism with focused Kerr magnetometry (whatever the laser power) in flakes, uncapped and deposited on SiO$_2$/Si, and thinner than 20~nm, were always unsuccessful. This is an indication that the measurement itself may heat the thin flakes and transform them, even at low laser power. Seeking further confirmation, we encapsulated a large flake, comprising extended (flat) regions of different thicknesses (5.0~nm, 6.9~nm, 15~nm, 25~nm, 50~nm, 100~nm), between two $h$-BN layers (135~nm and 5.2~nm). An optical micrograph is shown in Fig.~\ref{fig5}a. Everywhere on the sample, we only observe the two characteristic peaks of 1$T$-CrTe$_2$ in Raman scattering spectra (a representative spectrum, acquired on a 6.9~nm-thick region, is shown in Fig.~\ref{fig5}b). Strikingly, we measure, for these regions (and all other, thicker ones) a clear hysteretic signal in focused-Kerr magnetometry performed at room temperature (Fig.~\ref{fig5}c; the data, being very noisy, have been smoothened using Savitzky-Golay numerical post-filtering with a 31-point window and a third order polynomial), i.e. the flake is ferromagnetic.

The $h$-BN-capped 11-layer thick 1$T$-CrTe$_2$ is hence remarkably protected from phase transformation, being only marginally heated by the laser beam used for the measurements, and is ferromagnetic at room temperature. On the contrary, we could not detect any hysteretic signal for the thinner regions (5.0~nm, \textit{i.e.} 8~ML). Provided that our setup's sensitivity  is sufficient for such a small amount of magnetic matter, this suggests a reduced Curie temperature (the Raman spectra are indicative of unaltered 1$T$-CrTe$_2$), in other words $T_\mathrm{c}$ may become smaller than room temperature for a thickness between 8 and 11~ML. An alternative interpretation, not excluding the previous one, is a slight heating of the flake bringing it close to $T_\mathrm{c}$ (magnetization decreases very rapidly with temperature around $T_\mathrm{c}$, which is very close to 300~K).

\textit{\textbf{Conclusions and prospects. -- }}
We have shown that when thin flakes of the room-temperature ferromagnet 1$T$-CrTe$_2$ receive a heat load from a (focused) laser beam, they can locally and irreversibly transform into other Cr-Te compound, especially Cr$_5$Te$_8$. The effect can be controlled with the help of the substrate. The most common substrate for 2D or van der Waals materials, SiO$_2$/Si, acknowledgedly a poor heat conductor, limits heat dissipation from 1$T$-CrTe$_2$, which then transforms, at moderate laser powers such as those used in laser-based spectroscopy or magnetometry techniques. In contrast, good heat conductors, either metallic or insulating (Pt/Ta or $h$-BN), prevent laser-induced transformation within a 1~$\mu$m-wide 3~mW laser spot. The transformation of 1$T$-CrTe$_2$ into Cr-interleaved Cr-Te compounds such as Cr$_5$Te$_8$ is accompanied by a change of magnetic properties, of Curie temperature among other things, from above to below 300~K.

Exploiting the laser-induced transformation, we have drawn lateral magnetic patterns within 1$T$-CrTe$_2$. A first approach simply consisted in scanning a laser beam on 1$T$-CrTe$_2$ placed on a SiO$_2$/Si substrate. A second approach --- \textit{a priori} not limited in terms of spatial resolution by the diffraction limit --- was demonstrated, whereby a $h$-BN buffer layer between the flakes and SiO$_2$/Si was structured, leaving some 1$T$-CrTe$_2$ regions on $h$-BN and others directly on SiO$_2$/Si. Irradiating the full surface with a moderate laser power density then allowed to pattern 1$T$-CrTe$_2$.

Finally, we proposed $h$-BN capping as a protection, not against oxidation of the van der Waals materials as it is generally used for, but against heat-induced transformation. Doing so, we could observe room-temperature ferromagnetism in the thinnest 1$T$-CrTe$_2$ exfoliated so far, \textit{i.e.} only 11 layers.

A first implication of our findings is a caution warning for any laser-based characterization of 1$T$-CrTe$_2$ --- depending on the substrate, the lowest possible laser powers may be advisable. Capping with $h$-BN seems a relevant safety measure. On another note, and beyond the proof of principe we provided here, the second patterning approach we demonstrated holds potential for designing advanced spintronic structures. Historically, lateral spin valves, fabricated exclusively with metals (e.g. permalloy/Co/permalloy or permalloy/Ag/permalloy) brought new functionalities, e.g. giant magnetoresistance \cite{Jedema} and spin accumulation \cite{Kimura}. Hybrid systems combining a 2D material (graphene) with more standard materials (e.g. Co) already represented a new generation of lateral spin valves \cite{Tombros}; now fabrication approaches such as ours open the route to fully-van der Waals lateral spintronics systems.

For these systems, it will be crucial to control the width of the non-magnetic regions, for instance with electron-beam lithography of narrow patterns in a $h$-BN substrate. Much related is a key question we did not address, concerning lateral heat transport in 1$T$-CrTe$_2$ and how it will limit the narrowness of non-magnetic regions to be patterned within the 1$T$-CrTe$_2$ matrix.


\begin{acknowledgements}
We thank Michel Hehn for providing the Pt/Ta/Si substrates, as well as Alexandra Mougin and Jo\~{a}o Sampaio for useful comments on the manuscript and fruitful discussions. This work was supported by the Agence Nationale de la Recherche (ANR) through projects No. ANR-17-CE24-0007-03 `Bio-Ice’, ANR-19-CE24-0021 `ANETHUM’, ANR-23-CE09-0034 `NEXT', ANR-20-CE24-0017 `Matra2D', and under the program ESR/EquipEx+ (grant number ANR- 21-ESRE-0025). This work is also supported by France 2030 government investment plan managed by the French National Research Agency under grant reference PEPR SPIN – SPINMAT ANR-22-EXSP-0007. K.W. and T.T. acknowledge support from the JSPS KAKENHI (Grant Numbers 21H05233 and 23H02052) and World Premier International Research Center Initiative (WPI), MEXT, Japan.
\end{acknowledgements}

%

\end{document}